\documentstyle[sprocl,amssymb]{article}

 \newcommand{\bm}[1]{\mbox{\boldmath $#1$}}

 \newcommand{\zr}[1]{\mbox{\hspace*{#1em}}}

 \newcommand{\Id}{\mbox{1\zr{-0.62}{\small 1}}}
 \newcommand{\ZZ}{{\Bbb Z}}
 \newcommand{\NN}{{\Bbb N}}
 
 \newcommand{\RR}{{\Bbb R}}
 \newcommand{\QQ}{{\Bbb Q}}

 \newcommand{\wzwei}{\mbox{\footnotesize $\sqrt{2}\,$}}
 \newcommand{\wfuenf}{\mbox{\footnotesize $\sqrt{5}\,$}} 
 \newcommand{\mini}{\mbox{\tiny $\bullet$}}

 \newcommand{\aaa}{{\mathfrak a}}

\begin{document}

% \title{ }
\title{INVARIANT SUBMODULES AND SEMIGROUPS OF 
       SELF-SIMILARITIES FOR FIBONACCI MODULES}

\author{\sc Michael Baake {}\footnote{Heisenberg-fellow}}

\address{Institut f\"ur Theoretische Physik, Universit\"at T\"ubingen, \\
Auf der Morgenstelle 14, D-72076 T\"ubingen, Germany}

\author{\sc Robert V.~Moody}

\address{Department of Mathematical Sciences, \\ 
University of Alberta, Edmonton, Canada T6G 2G1}


\maketitle\abstracts{
The problem of invariance and self-similarity in $\ZZ$-modules is investigated.
For a selection
of examples relevant to quasicrystals, especially Fibonacci
modules, we determine the semigroup 
of self-similarities and encapsulate
the number of similarity submodules in terms of Dirichlet series generating 
functions. } 

\section{Introduction}

In the theory of quasicrystals, one typically deals with coordinates
(indices) that lie in some $\ZZ$-module of rank higher than
the dimension of the ambient space, $\RR^n$. Often, these $\ZZ$-modules
are $R$-modules for some ring $R$ and thereby admit sets of 
self-similarities (by scaling with elements of $R$), with a much richer
structure than that available for lattices in $\RR^n$. 
It is natural to study these and other
self-similarities, partly because they are intrinsic to the 
mathematics of the models in question, and partly to compensate
for the loss of translational symmetry in non-periodic structures.

In this contribution, we illustrate this concept with examples that 
show up in the context of quasicrystals and incommensurate structures. 
We shall focus here mainly on a class of modules with both crystallographic
symmetries and scaling invariance, especially by $\tau = (1 + \wfuenf)/2$.
We will describe both the semigroup of self-similarities and the set of
invariant submodules, encapsulating the number of submodules of given
index, $a(m)$, in terms of Dirichlet series generating functions.~\cite{Apostol}
They are more appropriate than power series because $a(m)$ will be a
multiplicative function, i.e.\ $a(1)=1$ and $a(mn) = a(m) a(n)$ for
$m,n$ coprime. Apart from providing a closed expression, such generating
functions also allow for the determination of asymptotic properties; see
Ref.~13 for some general background and Ref.~10 for some closely related
examples.

\section{Setup and general results}

We are interested in $\ZZ$-modules $M$ of finite rank $r$ that span
a  Euclidean 
space $\RR^n$, $r\geq n$, and the set of all affine {\em self-similarities}
\begin{equation}
   f \; : \quad x \; \mapsto \; \alpha R x + v
\end{equation}
that map $M$ into itself, where $R\in {\rm O}(n,\RR)$, 
$\alpha\in\RR_{}^{\mini}=\RR\backslash\{0\}$ 
(the inflation factor) \footnote{In general, if
$A$ is a ring, then $A_{}^{\mini}$ denotes its non-zero elements.}, 
and $v\in\RR^n$ (the translational part). 
Since $f(M)\subset M$, $f(0) = v \in M$
and $\alpha R(M) = f(M) - v \subset M$. The set of
all affine self-similarities of $M$ thus forms a {\em semigroup}, ${\cal S}(M)$,
which decomposes into a {\em linear} part, ${\cal L}(M)$ (those mappings fixing
the origin), and a {\em translational} part isomorphic to $(M,+)$:
\begin{equation}
      {\cal S}(M) \; = \; (M,+) \times_s {\cal L}(M) \, .
\end{equation}

Obviously, the main interest is in the subsemigroup ${\cal L}(M)$
which we will thus concentrate on in the sequel. We call
its (unique) maximal subgroup, denoted by ${\cal L}_1(M)$,
 the {\em group of units} of ${\cal L}(M)$,
and the modules $f(M)$ with $f\in{\cal L}(M)$ the {\em similarity submodules}
of $M$. They are {\em invariant} in the sense that they possess the same
point symmetry as $M$ (i.e.\ the point symmetry groups of $M$ and $f(M)$ are
conjugate in O$(n,\RR)$) and the same scaling invariance. 
In our setting, all such submodules are of finite index, and one is then
interested in the Dirichlet series generating function that counts them.
{}Following the practice of analytic number theory, this is called the 
{\em zeta-function} of $M$:
\begin{equation}
      \zeta^{}_M(s) \; := \; \sum_{m=1}^{\infty} \frac{a(m)}{m^s}
                    \; = \; \sum_{\tilde{M}} {1\over [M:\tilde{M}]^s_{}}
\end{equation}
where $\tilde{M}$ runs through all similarity submodules of $M$.
As mentioned above, $a(m)$ is an arithmetic function, and will be
multiplicative in all our examples. Note that $s$ is a complex number,
and that such series have a nice analytic behaviour, e.g.\ they converge
absolutely for ${\rm Re}(s) > c$ for some $c\in\RR\,$, see Ref.~1 for a more
general setting and further details.

In the happy instance that $M$ may actually be viewed as
the ring of integers of some number field $K$ with class
number $1$ (i.e.\ unique factorization), the scaling invariance
of $M$ implies that the similarity submodules coincide with the 
non-zero {\em ideals} of $M$ (i.e.\ subgroups $\aaa\subset M$, $\aaa\neq 0$,
with $M\aaa\subset\aaa$), ${\cal L}_1(M)$ is the 
group of {\em units} of $M$ (i.e.\ the group of invertible elements of $M$), and
$\zeta^{}_M$ is the {\em Dedekind zeta-function} of $K$ (see Ref~3., Thm.~2).
The simplest such case is $M=\ZZ$ where the non-zero ideals are
$m\ZZ$, $m\in\NN$, and the Dirichlet series for $M$ then is the well-known 
Riemann zeta-function~\cite{Apostol} itself,
$\zeta(s) = \sum_{m=1}^{\infty} m^{-s}$.

An interesting class to study is that of the
{\em Fibonacci modules}, i.e.\ of modules $M$ such that 
$x\cdot y \subset \ZZ[\tau]$ for all $x,y\in M$. Since some of them
have already been discussed elsewhere \cite{Baake,PBR,MB,BM},
we focus here on (hyper-)cubic $\ZZ[\tau]$-modules of the form 
$\ZZ[\tau]\bm{e}^{}_1 \oplus \ldots \oplus \ZZ[\tau]\bm{e}^{}_n$,
with $\{\bm{e}^{}_1, \ldots, \bm{e}^{}_n \}$ the standard
Euclidean basis of $\RR^n$. Note that they possess (hyper-)cubic
point symmetry because ${\rm O}(n,\ZZ[\tau]) = {\rm O}(n,\ZZ)
\simeq (C_2)^n \times_s S_n$, which is the hyperoctahedral group 
(of order $2^n n!$).  We also deal with 
$M=\ZZ[\wzwei]\bm{e}^{}_1 \oplus \ZZ[\wzwei]\bm{e}^{}_2$.

\section{A one-dimensional example}

The simplest Fibonacci module is the ring $\ZZ[\tau]$ itself.
It is, at the same time, a $\ZZ$-module of rank 2 and a 
$\ZZ[\tau]$-module of rank 1. Seen as the former, it contains
$\sigma^{}_1(m)$ $\ZZ$-submodules of index $m$, where 
$\sigma^{}_1(m) = \sum_{d|m} d$ is a divisor function, see
Appendix A of Ref.~2 for details. Though all these submodules
are invariant under inversion, most of them are {\em not}
invariant under multiplication by $\tau$. Since we also demand
that, we have to search for all $\ZZ[\tau]$-submodules of $\ZZ[\tau]$,
hence for all {\em ideals} $\aaa\subset\ZZ[\tau]$, where we exclude
the trivial case $\aaa=0$. 

Since $\ZZ[\tau]$
is the ring of integers in the quadratic field $\QQ(\tau)$, 
we are in the favorable situation mentioned above. 
Obviously, ${\cal L}(\ZZ[\tau]) = (\ZZ[\tau]_{}^{\mini}, \cdot)$, and the
group of units is ${\cal L}_1(\ZZ[\tau]) = \{ \pm \tau^m \mid m\in\ZZ\}
\simeq C_2\times C_{\infty}$. The Dedekind zeta-function 
of this field is the generating function we want. It reads
\begin{eqnarray} \label{tau-zeta}
\lefteqn{ \zeta_{\QQ(\tau)}^{}(s) \; = \; 
            \frac{1}{1-5^{-s}} \cdot
            \prod_{p \equiv \pm 1 \; (5)} 
                   \frac{1}{(1-p^{-s})^2} \cdot
            \prod_{p \equiv \pm 2 \; (5)} 
                   \frac{1}{1-p^{-2s}} } \\
&  & \!\!\!\! \mbox{\small $
 = \,  1+{1\over4^s}+{1\over5^s}+{1\over9^s}+{2\over11^s}+{1\over16^s}+
       {2\over19^s}+{1\over20^s}+{1\over25^s}+{2\over29^s}+{2\over31^s}+
       {1\over36^s}+ {2\over41^s} % +{2\over44^s}
     + \cdots $}  \nonumber
\end{eqnarray}
Here, we made use of another property of Dirichlet series of multiplicative
arithmetic functions, namely the existence of an Euler product 
representation~\cite{Apostol} where the product runs over all rational primes
with the restrictions indicated.
Note that this generating function, as in the examples below, allows for the 
determination of asymptotic properties, see Refs.\ 2, 3 and 10 for examples.

\section{Planar examples}

The most important planar examples are probably the modules with
$n$-fold symmetry, such as the rings of cyclotomic integers, 
which are discussed in detail in Refs.\ 3, 5 and 10. 
To complement this, let us consider the Fibonacci 
module ${\cal M} = \ZZ[\tau]\bm{e}^{}_1 \oplus \ZZ[\tau]\bm{e}^{}_2$ 
which has only the point symmetry of the square lattice, and would appear
in a $\tau$-modulated version of it.
 
In this section, we describe
the orientation preserving transformations only (indicating this by
a superscript $+$). To obtain all transformations, we form
the semi-direct product with another $C_2$ corresponding to complex conjugation.

We first observe that we may view $\cal M$ as a ring,
namely ${\cal M} = \ZZ[i \tau] = \ZZ[i,\tau]$.
It is not hard to show that $\cal M$ is the ring of integers
of the quartic field $K:=\QQ(i \tau)$, the splitting field of the
polynomial $x^4 + 3 x^2 + 1$. 
The Galois group of $K$ is $C_2\times C_2$ and $K$
is a totally imaginary extension of its
maximal real subfield, $\QQ(\tau) = \QQ(\wfuenf)$
(so, $K$ has degree 2 over $\QQ(\tau)$).
The latter has class number $h(\QQ(\tau))=1$, and hence unique 
factorization. From Ref.~6
(see pp.~11, 46, and the Table of relative class numbers), one can now 
calculate that the class number
of $\QQ(i \tau)$ is $h(\QQ(i \tau)) = h^* \cdot h(\QQ(\tau))$ where $h^*$,
the relative class number of $\QQ(i \tau)/\QQ(\tau)$, is 1. Hence,
$h(\QQ(i \tau))=1$, and we have unique factorization again. This
means that all ideals of $\ZZ[i \tau]$ are principal and thus
coincide with the similarity submodules. 
So, ${\cal L}^+({\cal M}) = ({\cal M}_{}^{\mini},\cdot)$ and
we now have to determine the group of units and the zeta-function of $K$. 
One can prove (though we have to omit that here) that the former is given by
\begin{equation}
       {\cal L}_1^+({\cal M}) \; = \; \{ i^k \tau^{\ell} \mid 
         k=0,1,2,3 \mbox{ and } \ell\in\ZZ \} 
        \; \simeq \; C_4 \times C_{\infty} \, .
\end{equation}

To calculate the Dirichlet series generating function of the number
of similarity submodules ($=$ ideals) of index $m$, one now realizes
that $\QQ(i \tau)$ is a degree 2 subfield of the 20th cyclotomic field,
$K_{20} = \QQ(i \xi)$, where $\xi = e^{2\pi i/5}$. Now, one can apply
the technique of Dirichlet characters, see chapters 3 and 4 of Ref.~12,
to determine the Dedekind zeta-function of $\QQ(i \tau)$
as a product of the $L$-functions of four characters. The result reads
\begin{eqnarray}
\lefteqn{ \zeta_{\cal M}^{}(s) \; = \; 
            \frac{1}{1-4^{-s}} \cdot \frac{1}{(1-5^{-s})^2} \cdot
            \prod_{p} 
                   \frac{1}{(1-p^{-s})^4} \cdot
            \prod_{\tilde{p}} 
                   \frac{1}{(1-{\tilde p}^{-2s})^2} } \\
&  & \!\!\!\! \mbox{\small $
 = \,  1+{1\over4^s}+{2\over5^s}+{2\over9^s}+{1\over16^s}+{2\over20^s}
        +{3\over25^s}+{2\over36^s}+{4\over45^s}+{2\over49^s}
        +{1\over64^s}+{2\over80^s}+{3\over81^s}
        + \cdots $}  \nonumber
\end{eqnarray}
where $p$ runs over primes $\equiv 1,9 \; (20)$ and $\tilde{p}$
over those $\equiv 3,7,11,13,17,19 \; (20)$.

Before we continue with the Fibonacci modules, let us briefly mention
another example, $M = \ZZ[\wzwei]\bm{e}^{}_1 \oplus \ZZ[\wzwei]\bm{e}^{}_2
= \ZZ^2 \oplus \wzwei \cdot \ZZ^2$.
With $\xi = e^{2\pi i/8}$, one sees that 
$M$ is the ring $\ZZ[i,\wzwei] \subset \ZZ[\xi]$, 
and the field generated by $M$ is $\QQ(\xi)$. This time, although
$\QQ(\xi)$ has class number $1$, $M$ lies with index $2$ in its ring of 
integers, $\ZZ[\xi]$. In fact, $M$ is not integrally closed, nor is
it a principal ideal domain. However, 
${\cal L}^+(M) = (M_{}^{\mini},\cdot)$, and the group of units,
since $\xi\not\in M$, turns out to be (with $\lambda = 1 + \wzwei$)
\begin{equation}
     {\cal L}_1^+(M) \; = \; \{ i^k \lambda^{\ell} \mid
       k = 0,1,2,3 \;\; {\rm and} \;\; \ell\in\ZZ \}
       \; \simeq \; C_4 \times C_{\infty} \, .
\end{equation}

Our further analysis revolves around the prime $2$, which
splits (ignoring units) as $(1+\xi)^4$. However, the prime
$(1+\xi)$ itself does not occur in $M$, and, in fact, $M\cap (1+\xi)\ZZ[\xi]
=(1+\xi)^2\ZZ[\xi] = (1+ i)\ZZ[\xi]$. Not all ideals of $M$ are principal,
but only these are similarity submodules.
There is a one-to-one correspondence between odd principal ideals
of $M$ (i.e.\ those of odd index) and those of $\ZZ[\xi]$, given by 
$aM \leftrightarrow a\ZZ[\xi],~ a\in M$. The even principal ideals of $M$
require some care. There are none of index $2(2\ell+1)$, but twice as many
of any other even index than there are in $\ZZ[\xi]$.
In fact, if $\aaa = a M$ and $(1+i)\mid a$, then $\tilde{\aaa} = \xi a M$
is another ideal, but both are inside $a\ZZ[\xi]=\xi a\ZZ[\xi]$ (this is
due to $\xi$ being a unit of $\ZZ[\xi]$, but not of $M$).
Now, we have $[M:aM] = [\ZZ[\xi]:a\ZZ[\xi]]$
for all $0 \neq a \in M$ (note that for $a \in M$, $aM$
is never an ideal of $\ZZ[\xi]$).

So, for counting the principal ideals of $M$, we have to remove 
from the zeta function of $\QQ(\xi)$
those terms that count
ideals $\aaa$ for which $N(\aaa)$ has the form $2(2l +1)$, and to 
count all other even ideals twice. We thus obtain:
\begin{eqnarray}
   \lefteqn{ \zeta^{}_M(s) \; = \; (1 - 2^{-s} + 2 \cdot 4^{-s}) 
                          \cdot \zeta^{}_{\QQ(\xi)}(s) }  \\
&  & \!\!\!\! \mbox{\small $
 = \, 1+{2\over4^s}+{2\over8^s}+{2\over9^s}+{2\over16^s}
       +{4\over17^s}+{2\over25^s}+{2\over32^s}+{4\over36^s}
       +{4\over41^s}+{2\over49^s}+{2\over64^s}+{8\over68^s} 
       + \ldots $}  \nonumber
\end{eqnarray}
where $\zeta^{}_{\QQ(\xi)}(s)$ is given explicitly in Ref.~3.

\section{An example in three dimensions}

Let us finally treat the module ${\cal M}_c = \ZZ[\tau]^3$ 
which is closely related to the icosahedral modules \cite{Baake,BM}. 
To do so, we have to return to the original definition
of (linear) similarities, i.e.\ to $f(x)=\alpha R x$ and
the condition $f({\cal M}_c)\subset {\cal M}_c$. 
In terms of the basis $\{\bm{e}^{}_1, \ldots, \bm{e}^{}_3\}$, all entries of the
matrix $\alpha R$ must be integral, i.e.\ in $\ZZ[\tau]$.
With $\alpha R$, also $\alpha R^t$ must be integral, and orthogonality of 
$R$ implies $(\alpha R)(\alpha R^t) = \alpha^2 \Id\,$, hence $\alpha^2\in\ZZ[\tau]$.
Together with $\det(\alpha R) = \alpha^3 \in\ZZ[\tau]$, 
this gives $\alpha\in\QQ(\tau)$,
and $\alpha^2\in\ZZ[\tau]$ is then only possible for $\alpha\in\ZZ[\tau]$.

With this result, $\alpha R$ integral means $R\in {\rm O}(3,\QQ(\tau))$
and $\alpha\in\ZZ[\tau]$ must be a $\ZZ[\tau]$-multiple of the
denominator of $R$,
\begin{equation}
    {\rm den}(R) \; := \; \gcd\{\beta\in\ZZ[\tau]_+ \mid
                                \beta R \;\; {\rm integral}\, \}\, ,
\end{equation} 
which is well defined, even as a number rather than an ideal,
up to units, i.e.\ up to $\pm \tau^m$, $m\in\ZZ$, due to
unique factorization in $\ZZ[\tau]$. Note that
$\pm \tau^m {\cal M}_c = {\cal M}_c$.

With ${\cal R} := \{{\rm den}(R)\cdot R\mid R\in {\rm O}(3,\QQ(\tau))\}$,
we characterize ${\cal L}({\cal M}_c)$ as 
\begin{equation}
    {\cal L}({\cal M}_c) \; = \; (\{\ZZ[\tau]_{}^{\mini} \times {\cal R}\}, \cdot)
\end{equation}
and the group of units in it turns out to be
\begin{equation}
    {\cal L}_1({\cal M}_c)
       \; = \; \{\tau^m\mid m\in\ZZ\} \times {\rm O}(3,\ZZ)
       \; \simeq \; C_{\infty} \times ((C_2)^3 \times_s S_3) \, .
\end{equation}
Note that an analogous result holds in general for 
(hyper-)cubic Fibonacci modules
in Euclidean spaces of odd dimension.

{}Finally, we have to count the similarity submodules of ${\cal M}_c$.
For a mapping $f(x)=\alpha\cdot{\rm den}(R)\cdot R x$, $\alpha\in\ZZ[\tau]$,
we have the index formula
\begin{equation}
   {\rm ind}(f) \; = \; [{\cal M}_c : f({\cal M}_c)]
       \; = \; | {\rm N}[\alpha]^3 \cdot 
                 {\rm N}[{\rm den}(R)]^3 | \, .
\end{equation}
The solution of the coincidence problem \cite{MB}
gives us the generating function for $1/24$ times the number
of SO$(3,\QQ(\tau))$ matrices $R$ with
$|{\rm N}[{\rm den}(R)]| = m$, 
\begin{equation}
   \Phi_c(s) \; = \; \frac{1+4^{1-s}}{1+4^{-s}} \cdot
      \frac{\zeta^{}_{\QQ(\tau)}(s)\zeta^{}_{\QQ(\tau)}(s-1)}
           {\zeta^{}_{\QQ(\tau)}(2s)} \, ,
\end{equation}
with the zeta-function $\zeta^{}_{\QQ(\tau)}(s)$ as given in Eq.~(\ref{tau-zeta}).

But then, since precisely 24 different SO$(3,\QQ(\tau))$ matrices
give rise to the same similarity submodule,
we finally get the following Dirichlet series generating function
for the number of similarity submodules of given index
\begin{eqnarray}
\lefteqn{F_{{\cal M}_c} (s) \; = \; 
      \zeta^{}_{\QQ(\tau)}(3s) \Phi_c(3s)   } \\
&  & \! \mbox{\small $
 = \,  1+{9\over64^s}+{7\over125^s}+{11\over729^s}+{26\over1331^s}
        +{41\over4096^s}+{42\over6859^s}+{63\over8000^s}
        +{37\over15625^s}+{62\over24389^s}
        + \cdots $}  \nonumber
\end{eqnarray}
Here, the factor $\zeta^{}_{\QQ(\tau)}(3s)$ takes care of the freedom
to go from a maximal submodule $M$ to a scaled copy, $\alpha M$,
with $\alpha\in\ZZ[\tau]$. Note that similar arguments apply to
$\ZZ[\tau]$-extensions of face-centred and body-centred cubic lattices.

\section{Concluding remarks}

With the methods and examples presented here, which apply mainly
to modulated structures, and with the explicit results of Refs.~3 and 5,
the structure of invariant submodules and their semigroups
of self-similarities should be transparent enough so that other examples
can be worked out along the same lines.
 
Among the applications is the determination of the possible colourings
of a given module if one demands that one colour occupies an invariant
submodule and the others code the cosets, see Refs.~3,4,7,8,11 for examples.

An extension to four dimensions, covering the 4D cubic lattices and the
Elser-Sloane quasicrystal, is also possible due to their relation to maximal
orders in the skew-field of quaternions, such as Hurwitz' ring of integral 
quaternions or the icosian ring \cite{Moody,MB}. This will be reported separately.

\section*{Acknowledgements}
We are grateful to Peter A.\thinspace B.\thinspace Pleasants for 
his most helpful comments in the preparation of this paper. 
This work was supported by the Volkswagen Stiftung
(RiP-program at Oberwolfach).


\section*{References}

\begin{thebibliography}{99}
% \small

\bibitem{Apostol}
T.~M.~Apostol, {\em Introduction to Analytic Number Theory},
Springer, New York (1976).

\bibitem{MB}
M.~Baake, ``Solution of the coincidence problem in dimensions $d\leq 4$'',
in: {\em The Mathematics of Long-Range Aperiodic Order}, ed.\ R.~V.~Moody, 
NATO ASI Series C 489, Kluwer, Dordrecht (1997), pp.\ 9--44.

\bibitem{Baake}
M.~Baake, ``Combinatorial aspects of colour symmetries'',
{\it J.\ Phys.} {\bf A 30} (1997) 2687--98.

\bibitem{BHLS}
M.~Baake, J.~Hermisson, R.~L\"uck and M.~Scheffer,
``Colourings of quasicrystals'',
in: {\em Proceedings of ICQ6}, ed.\ T.\ Fujiwara,
World Scientific, Singapore (1997), in press. 

\bibitem{BM}
M.~Baake and R.~V.~Moody, ``Similarity submodules and semigroups'',
in: {\em Quasicrystals and Discrete Geometry}, ed.\ J.~Patera, Fields 
Institute Publications, AMS, Rhode Island (1997), in press.

\bibitem{Hasse}
H.~Hasse, {\it \"Uber die Klassenzahl abelscher Zahlk\"orper},
Akademie, Berlin (1952).

\bibitem{Lifshitz}
R.~Lifshitz, ``Lattice color groups of quasicrystals'',
in: {\em Proceedings of ICQ6}, ed.\ T.\ Fujiwara,
World Scientific, Singapore (1997), in press; and:
``Theory of color symmetry for periodic and quasiperiodic crystals'', 
{\it Rev.\ Mod.\ Phys.} {\bf 69} (1997), in press.

\bibitem{Lueck}
R. L\"uck, ``Farbsymmetrien in kolorierten Fibonaccifolgen'',
DPG Fr\"uh\-jahrs\-tagung, Berlin (1995), 
{\em Phys.\ Verhandl.} {\bf 7} (1995) 1433.

\bibitem{Moody}
R.~V.~Moody and J.~Patera, ``Quasicrystals and icosians'',
{\em J.\ Phys.} {\bf A 26} (1993) 2829--53.

\bibitem{PBR}
P.~A.~B.~Pleasants, M.~Baake and J.~Roth, ``Planar coincidences with
$N$-fold symmetry'', {\it J.\ Math.\ Phys.} {\bf 37} (1996) 1029--58.

\bibitem{Max}
M.~Scheffer, PhD thesis, Univ.\ Stuttgart (1997), in preparation.

\bibitem{Wash}
L.~C.~Washington, {\it Introduction to Cyclotomic Fields},
2nd ed., Springer, New York (1997).

\bibitem{Wilf}
H.~S.~Wilf, {\em Generatingfunctionology}, 2nd ed., Academic Press,
Boston (1994).


\end{thebibliography}
\end{document}